\RequirePackage{fix-cm}
\documentclass[smallextended]{svjour3}       
\smartqed
\usepackage{graphicx}
\usepackage{amsmath}
\usepackage{floatflt}
\usepackage{amsfonts}
\usepackage{amssymb}
\usepackage{graphicx}
\usepackage{psfrag}
\usepackage{epstopdf}
\usepackage{slashed}
\usepackage{booktabs}
\usepackage{cite}
\usepackage{mathptmx}
\usepackage{latexsym}
\usepackage{chemarrow}
\usepackage{threeparttable}
\usepackage{array}
\usepackage{multirow}
\usepackage{lineno}
\usepackage[english]{babel}
\usepackage[latin1]{inputenc}
\usepackage[colorlinks,linkcolor=blue,anchorcolor=blue,citecolor=blue]{hyperref}
\usepackage{amssymb,xcolor}
\usepackage{graphicx}
\usepackage{exscale}
\usepackage{latexsym}
\usepackage{cases}
\usepackage{epstopdf}
\usepackage{subfigure}
\usepackage{verbatim}
\usepackage{multirow}
\usepackage{float}
\usepackage{bm}
\usepackage{caption}
\usepackage{color}

\newcommand{\ket}[1]{\left|#1\right\rangle}
\newcommand{\bra}[1]{\left\langle #1\right|}
\newcommand{\Tr}{\mathrm{Tr}}

\newcommand{\n}{\nonumber\\}

\newcommand{\Ss}{\mathbf{S}}

\newcommand{\Up}{\Uparrow}
\newcommand{\up}{\uparrow}
\newcommand{\Down}{\Downarrow}
\newcommand{\down}{\downarrow}
\newcommand{\ex}[1]{\langle #1\rangle}
\newcommand{\lf}{\left}
\newcommand{\rg}{\right}
\begin{document}

\nolinenumbers
\title{\textbf{Decoherence dynamics of entangled quantum states in the $XXX$ central spin model}}

\author{Qing-Kun Wan\and Hai-Long Shi\and  Xu Zhou \and Xiao-Hui Wang\and Wen-Li Yang}
\institute{Qing-Kun Wan\and Xu Zhou\and Wen-Li Yang\at
              Institute of Modern Physics, Northwest University, Xi'an 720127, China\\
              \email{xhwang@nwu.edu.cn}
           \and
              Qing-Kun Wan\and  Xu Zhou \and Xiao-Hui Wang\and Wen-Li Yang \at
              School of Physics, Northwest University, Xi'an 720127, China
           \and
              Qing-Kun Wan\and Hai-Long Shi\at
              State Key Laboratory of Magnetic Resonance and Atomic and Molecular Physics, Wuhan Institute of Physics and Mathematics, Chinese Academy of Sciences, Wuhan 430071, China
           \and
            Qing-Kun Wan\and Hai-Long Shi\at
              University of Chinese Academy of Sciences, Beijing 100049, China
            \and
            Xiao-Hui Wang\and Wen-Li Yang\at
              Shaanxi Key Laboratory for Theoretical Physics Frontiers, Xi'an 720127, China
}

\date{Received: date / Accepted: date}
\maketitle
\begin{abstract}
Maintaining  coherence of a qubit is of vital importance for realizing a large-scale quantum computer in practice.
In this work, we study the central spin decoherence problem in the $XXX$ central spin model (CSM) and focus on the quantum states with different initial entanglement, namely intra-bath entanglement or system-bath entanglement.
We analytically obtain  their evolutions of fidelity, entanglement, and quantum coherence.
When the initial bath spins constitute an $N$-particle entangled state (the Greenberger-Horne-Zeilinger-bath or the $W$-bath), the leading  amplitudes of their fidelity evolutions both scale as  $\mathcal O(1/N)$,
which is the same as the case of a fully polarized bath.
However, when the central spin is maximally entangled with one of the bath spins, the amplitude scaling of its fidelity evolution declines from $\mathcal O(1/N)$ to $\mathcal O(1/N^2)$.
That implies appropriate initial system-bath entanglement is contributive to suppress central spin decoherence.
In addition, with the help of system-bath entanglement, we realize quantum coherence-enhanced dynamics for the central spin where the consumption of bath entanglement is shown to play a central role.

\end{abstract}

\section{Introduction}
A spin of a localized electron in semiconductor quantum dots (QDs) is a promising candidate  for a physical matter qubit---the elementary unit of a quantum computer \cite{Imamoglu1,Oestreich,Imamoglu2,Pla,Loss,Veldhorst}.
A key challenge of realizing solid-state quantum computation is suppressing electron spin relaxation, namely decoherence, so that 
quantum information can be stored and
manipulated without loss in a sufficiently long coherence time.
Decoherence of an electron spin is induced by the inevitable hyperfine interaction with the surrounding nuclei \cite{Merkulov,Khaetskii1,Semenov,Braun}, which is captured by  the Hamiltonian of the central spin model (CSM) as follows:
\begin{equation}\label{inh-CSM}
H=\sum_{j=1}^NA_j\mathbf{S}_0\cdot\mathbf{S}_j,
\end{equation} 
where a central spin $\mathbf{S_0}$ is coupled to a spin bath of $N$ nuclei $\mathbf{S}_j$ via an inhomogeneous hyperfine interaction $A_j$. 
Increasing attention has been paid to this model for seeking theoretical guidance of suppressing central spin decoherence \cite{Claeys,Guan,Seo,Liu,Wu,Anders}.

A great deal of investigations to the decoherence problem were mainly focused on some special initial product states. 
For the initial state with a fully polarized bath, the central spin polarization function $\ex{S_0^z(t)}$ quantifying the degree of amplitude decoherence  oscillates with a frequency $\sim\sum_{j=1}^N A_j$ and an amplitude of order $\mathcal O(1/N)$ about a mean value \cite{Bortz1,Bortz2} while for an unpolarized bath as the initial bath the corresponding oscillation with a frequency $\mathcal O(\sqrt{N})$ and an amplitude $\mathcal O(1)$ \cite{Khaetskii1,Khaetskii}.
This observation indicates that the decay of the central spin can be suppressed through polarized nuclear spins \cite{Hanson,Burkard}.
Recently, Floquet resonances have been suggested to realize such a polarization-based decoupling of the central spin from its environment in the CSM \cite{Claeys}.

Entanglement, as a fundamental quantum resource, takes responsibility for most quantum information processes \cite{QT-of-En,QT-En-2}. 
A paradigmatic example is the quantum teleportation where the use of maximally entangled states ensures the success of deterministic remote quantum state transfer \cite{tele}.
The implementation of this quantum technology in quantum dot chains or spin chains has been verified to be possible beyond the classical teleportation scheme \cite{dots chain,spin chain}.
This hints that quantum information can be protected with the entanglement generated by the spin chains.  
A natural question arises whether entanglement contained in initial states can protect the central spin from decoherence.
It was shown that decoherence of the central spin can be suppressed by using persistent entanglement in the bath \cite{Dawson}. 
And Ref. \cite{Su-Zhao} identified a coherence-preserving phase by the evolution of the bath concurrence.
There is, however, still a lack of further understanding of the role of initial entanglement played in the decoherence problem, especially for initial system-bath entanglement.

The inhomogeneous CSM (\ref{inh-CSM}) is exactly solvable by the Bethe ansatz. \cite{Gaudin, Richardson,Dukelsky}.
However, the difficulty of solving the Bethe ansatz equations prohibits full analytical access to the evolutions of quantum dynamics.
We bypass this obstacle by considering the homogeneous CSM in this paper.
The dynamics in the homogeneous model can be analytically calculated while providing some valid insights for the inhomogeneous CSM \cite{Bortz2,Barnes}.
In Sect. \ref{2}, we briefly review the homogeneous CSM and its exact solutions.
Some coherence and entanglement measures, such as fidelity, the relative entropy of coherence, and concurrence, are also introduced.
In Sect. \ref{decoherence-problem}, we obtain the  evolutions of these quantities for initial states with different types of entanglement, namely,
intra-bath entanglement [Greenberger-Horne-Zeilinger (GHZ) state as bath or the $W$ state as bath] or  system-bath entanglement. (The central spin and one of the bath spins form a maximal entangled pair.)
The initial state without entanglement, i.e., a product state with a fully polarized bath, is also considered for comparison.
It is observed that the amplitude scaling of fidelity reduces from $\mathcal O(1/N)$ to $\mathcal O(1/N^2)$ when the product state is replaced by the system-bath entangled state.
However, amplitude scalings of fidelity are not reduced to less than $\mathcal O(1/N)$  by the initial states with  GHZ-type or $W$-type bath entanglement. 
This demonstrates the importance of system-bath entanglement in suppression of decoherence.
In quantum resource theories, quantum coherence has been shown as a key resource \cite{Streltsov} to implement some quantum technologies, such as quantum channel discrimination \cite{Napoli,Piani} and quantum algorithms \cite{Shi,Anand,Matera}, by quantifying it with the relative entropy function.
Moreover, much effort has been devoted to investigate manipulation and distillation of quantum coherence within this framework \cite{Bai,Chitambar,Shi2,C2,Wu2}. 
Based on this consideration, the relative entropy of coherence will be adopted to study  quantum coherence dynamics of CSM   in Sect. \ref{Example}.
Eventually, we realize coherence-enhanced dynamics for some initial states with suitable system-bath entanglement where the consumption of entanglement in entangled pair is emphasized to explain the increase of quantum coherence in the central spin.
A summary is made in Sect. \ref{4}.

\section{The central spin model and quantum correlations}\label{2}
We consider a single electron confined in a quantum dot in which decoherence of the electron is induced by the homogeneous hyperfine interaction with  surrounding nuclei.
Setting $A_j=2$ in the Hamiltonian (\ref{inh-CSM}), we get
\begin{eqnarray}\label{Hamiltonian}
H=2\sum_{j=1}^N \Ss_0\cdot\Ss_j,
\end{eqnarray}
where $S_0^\alpha$ and $S_j^\alpha$ $(\alpha=x,y,z)$ denote spin-1/2 operators of the central spin and the $j$-th bath spin respectively.
This model is a simplified  Gaudin model, yet it still exhibits rich phenomena and is an ideal model for analytically investigating the decoherence problem.
By introducing a bath spin operator $\Ss_b=\sum_{j=1}^{N}\Ss_j$ and a total spin operator $\Ss_{tot}=\Ss_b+\Ss_0$, the Hamiltonian can be rewritten as  
\begin{equation}\label{CSM}
H=\Ss_{tot}^2-\Ss^2_b-\Ss^2_0,
\end{equation}

For a given initial state $\ket{\Psi(0)}$,  our  goal is to obtain the wave function under a unitary time evolution of the Hamiltonian (\ref{Hamiltonian}), i.e., $\ket{\Psi(t)}=\exp(-iHt)\ket{\Psi(0)}$,
then reduce the density matrix $\rho(t)=\ket{\Psi(t)}\bra{\Psi(t)}$ to some specified lattice sites,  and eventually use these reduced density matrices to 
calculate fidelity, concurrence, and the relative entropy of coherence.
For convenience, we introduce the following notation:
\begin{equation}\label{notation}
\ket{s_{[L]},s_{[L-1]},\ldots, s_{[M]},s_{[L]}^z}_{[L]},\quad 1\leq M\leq L,
\end{equation}
to denote a $L$-qubit state  which is the eigenstate of  $\mathbf{S}_{[K]}^2=(\sum_{j=L-K+1}^L\mathbf{S}_j)^2$ and $\mathbf{S}_{[K]}^z=\sum_{j=L-K+1}^L\mathbf{S}_j^z$ with eigenvalues $s_{[K]}(s_{[K]}+1)$ and  $s_{[K]}^z$, respectively.
For instance, the $N$-qubit GHZ state \cite{GHZ} $\ket{\rm GHZ}_{[N]}=( \ket{\up}^{\otimes N}+\ket{\downarrow}^{\otimes N} )/\sqrt{2}$ can be rewritten as $(\ket{N/2,N/2}_{[N]}+\ket{N/2,-N/2}_{[N]})/\sqrt{2}$.
When $L=N+1$ the quantum states (\ref{notation}) are exactly the eigenstates of the Hamiltonian (\ref{Hamiltonian}), i.e,
	\begin{eqnarray}\label{eigen-solution}
	&&H\ket{s_b-1/2,s_b,s^z}_{[N+1]}=(-s_b-1)\ket{s_b-1/2,s_b,s^z}_{[N+1]},\n
	&&H\ket{s_b+1/2,s_b,s^z}_{[N+1]}=s_b\ket{s_b+1/2,s_b,s^z}_{[N+1]},
	\end{eqnarray}
where $s_b=N/2,N/2-1,\ldots,0$ for $N$ being even and $s_b=N/2,N/2-1,\ldots,1/2$ for $N$ being odd.

The calculation of $\ket{\Psi(t)}$ is easy to carry out once we decompose $\ket{\Psi(0)}$ into the eigenstates of the Hamiltonian (\ref{Hamiltonian}).
This decomposition process can be implemented by repeatedly using the following relations:
	\begin{small}
		\begin{eqnarray}\label{eigen1}
		&&\ket{\downarrow}\ket{s_{[L]},s_{[L]}^z}_{[L]}=\n
		&&\frac{1}{\sqrt{2s_{[L]}\!+\!1}}\left(\sqrt{s_{[L]}\!-\!s_{[L]}^z\!+\!1}\ket{s_{[L]}\!+\!\frac{1}{2},s_{[L]},s_{[L]}^z\!-\!\frac{1}{2}}_{[L\!+\!1]}\!-\!\sqrt{s_{[L]}\!+\!s_{[L]}^z}\ket{s_{[L]}\!-\!\frac{1}{2},s_{[L]},s_{[L]}^z\!-\!\frac{1}{2}}_{[L\!+\!1]} \right),\n
		&&\ket{\up}\ket{s_{[L]},s_{[L]}^z\!-\!1}_{[L]}=\n
		&&\frac{1}{\sqrt{2s_{[L]}\!+\!1}}\left(\sqrt{s_{[L]}\!+\!s_{[L]}^z}\ket{s_{[L]}\!+\!\frac{1}{2},s_{[L]},s_{[L]}^z\!-\!\frac{1}{2}}_{[L\!+\!1]}\!+\!\sqrt{s_{[L]}\!-\!s_{[L]\!+\!1}^z\!+\!1}\ket{s_{[L]}\!-\!\frac{1}{2},s_{[L]},s_{[L]}^z\!-\!\frac{1}{2}}_{[L\!+\!1]} \right).\n
		\end{eqnarray}
	\end{small}More details can be found in Ref. \cite{Bortz2}.
	Note that the expression of eigenstates (\ref{eigen-solution}) is inconvenient to  evaluate the reduced density matrices of $\ket{\Psi(t)}$. 
	Thus, one needs to rewrite the $\ket{\Psi(t)}$ into a more explicit form by using the inverse relation of (\ref{eigen1}):
		\begin{eqnarray}\label{eigen-back}
		&&\ket{s_{[L]}\!+\!\frac{1}{2},s_{[L]},s_{[L]}^z\!-\!\frac{1}{2}}_{[L\!+\!1]}=\n
		&&\frac{1}{\sqrt{2s_{[L]}\!+\!1}}\left( \sqrt{s_{[L]}\!+\!s_{[L]}^z}\ket{\uparrow}\ket{s_{[L]},s_{[L]}^z\!-\!1}_{[L]}\!+\!\sqrt{s_{[L]}\!-\!s_{[L]}^z\!+\!1}\ket{\down}\ket{s_{[L]},s_{[L]}^z}_{[L]}\right),\n
		&&\ket{s_{[L]}\!-\!\frac{1}{2},s_{[L]},s_{[L]}^z\!-\!\frac{1}{2}}_{[L\!+\!1]}=\n
		&&\frac{1}{\sqrt{2s_{[L]}\!+\!1}}\left( \sqrt{s_{[L]}\!-\!s_{[L]}^z\!+\!1}\ket{\uparrow}\ket{s_{[L]},s_{[L]}^z\!-\!1}_{[L]}\!-\!\sqrt{s_{[L]}\!+\!s_{[L]}^z}\ket{\downarrow}\ket{s_{[L]},s_{[L]}^z}_{[L]}\right).\n
		\end{eqnarray}
Concrete examples are given in the next section.

A measure of distance between two quantum states is necessary to quantitatively characterize the degree of central spin decoherence. 
A widely used one is  fidelity which is defined to be \cite{Jozsa}
\begin{equation}\label{fidelity}
F_0(t)=\left(\mathrm{Tr}\sqrt{\rho_{0}(0)^{1/2}\rho_{0}(t)\rho_{0}(0)^{1/2}}\right) ^2,
\end{equation}
where $\rho_{0}(0)$ denotes the initial reduced density matrix of the central spin  and $\rho_0(t) =\Tr_{\rm bath} (\exp(-iHt)\rho(0)\exp(iHt))$.
Fidelity is bounded between 0 and 1.
If $\rho_{0}(0)$  is the same as $\rho_{0}(t)$ then the fidelity equals to one,
whereas if $\rho_0(0)$ is different from $\rho_0(t)$ then the fidelity is strictly less than one. 
Particularly, when $\rho_0(0)$ and $\rho_0(t)$ are perfectly distinguishable, i.e., they are supported on orthogonal subspaces, the fidelity reaches the minimal value zero \cite{Nielsen}.
As a consequence, the smaller fidelity indicates the central spin is more easily decoherent.

In a broad context, decoherence refers to the changes in the reduced density matrix of the central spin, including  amplitude decoherence and phase decoherence.
The former focuses on the changes of the  diagonal elements,
while the latter focuses on the non-diagonal elements.
These two different decoherence can be characterized via  $\ex{S^z_0(t)}$ or $\ex{S_0^+(t)}$ for some special initial states. 
In Sect. \ref{decoherence-problem}, we will consider the initial states that only suffers from the amplitude decoherence and use the fidelity to characterize it.
For phase decoherence problem, see Sect. \ref{Example}, 
we choose the relative entropy of coherence to evaluate the quantum coherence encoded in the central spin rather than $\ex{S_0^+(t)}$.
The reason lies in the fact that the quantum coherence characterized by the relative entropy has been shown to be a quantum resource and plays an indispensable role in quantum information science.
On the other hand, this coherence measure also satisfies the requirement of investigating the phase decoherence problem.
Now we explain it in more detail. 
In the field of quantum information, quantum coherence is unambiguously defined and has an operational property in which the states  without off-diagonal elements in their density matrices are incoherent states and the incoherent operations $\Lambda_{\rm ICPTP}$ are defined to be completely positive and trace preserving  quantum operations mapping incoherent states into incoherent states.
By attaching other reasonable requirements to coherence measures $\mathcal C$, e.g., $\mathcal C(\rho)\geq\mathcal C(\Lambda_{\rm ICPTP}(\rho))$, Ref. \cite{Baumgratz} established a rigorous mathematical framework for quantifying quantum coherence
where the relative entropy of coherence is an excellent measure \cite{Baumgratz}
\begin{equation}\label{relative-entropy}
C_0(t)=S\lf(\rho_0^D(t)\rg)-S\lf(\rho_0(t)\rg).
\end{equation}
Here, $S(\sigma)=-\Tr\left[ \sigma\ln(\sigma)\right] $ is the von Neumann entropy and $\rho_0^D(t)$ is obtained by deleting all off-diagonal elements in the reduced density matrix of the central spin.  

The effects of bipartite entanglement on CSM dynamics can be elucidated via the concurrence.
For a pure state $\ket{\psi}_{01}$, it is defined as
\cite{Wootters}
\begin{equation}\label{concurrence1}
E_{01}(\ket{\psi}_{01})=\sqrt{2(1-\gamma(\rho_0))},
\end{equation}
where $\gamma(\rho_0)=\Tr(\rho_0^2)$ is the purity of $\rho_0=\Tr_1(\ket{\psi}_{01}\bra{\psi}_{01})$.
For a mixed state $\rho_{01}$, one defines the concurrence as \cite{Wootters}
\begin{equation}\label{concurrence2}
E_{01}(\rho_{01})=\min_{\ket{\psi_j}}\sum_{j}p_j E_{01}(\ket{\psi_j}),
\end{equation}
where the minimization is over all ensemble decompositions $\rho_{01}=\sum_jp_j\ket{\psi_j}\bra{\psi_j}$.
We will use concurrence to calculate  the evolution of bipartite entanglement between the central spin and the first bath spin $\rho_{01}(t)$.
The closed form of the concurrence for two-qubit state $\rho_{01}(t)$ is  \cite{Wootters}
\begin{equation}\label{concurrence}
E_{01}(t)=\max\left\lbrace 0,\lambda_1(t)-\lambda_2(t)-\lambda_3(t)-\lambda_4(t)\right\rbrace,
\end{equation}
where $\lambda_1(t)$, $\lambda_2(t)$, $\lambda_3(t)$, $\lambda_4(t)$ are the square roots of the eigenvalues of $\rho_{01}(t)\tilde{\rho}_{01}(t)$ satisfying $\lambda_1(t)\geq\lambda_2(t)\geq\lambda_3(t)\geq\lambda_4(t)$. $\rho_{01}(t)$ is the reduced density matrix of the spin $0$ and $1$ by tracing out  other spins  and $\tilde{\rho}_{01}(t)=\sigma_0^y\otimes\sigma_1^y\rho_{01}^*(t)\sigma_0^y\otimes\sigma_1^y$.

\section{Decoherence problem}\label{decoherence-problem}
\subsection{The product bath}
Before discussing entangled baths, we first consider an initial state with a product bath given by 
\begin{equation}\label{Product-bath}
\ket{\Psi_P(0)}=\ket{\Downarrow}\ket{\uparrow\uparrow\cdots\uparrow}_{[N]},
\end{equation}
where the central spin is spin-down denoted by $\ket{\Down}$ and $N$ bath spins constitute  a fully polarized bath denoted by $\ket{\up\up\cdots\up}_{[N]}$.
This quantum state can be rewritten as $\ket{\Down}\ket{N/2,(N-1)/2,N/2}_{[N]}$ in our notation (\ref{notation}), and then by Eq.  (\ref{eigen1}) we obtain the  
state after a unitary time evolution as follows
	\begin{eqnarray}\label{Product-2}
	\ket{\Psi_P(t)}&=&\frac{1}{\sqrt{N+1}}e^{-i\frac{N}{2}t}\ket{\frac{N+1}{2},\frac{N}{2},\frac{N-1}{2},\frac{N-1}{2}}_{[N+1]}\n
	& &-\sqrt{\frac{N}{N+1}}e^{i\frac{N+2}{2}t}\ket{\frac{N-1}{2},\frac{N}{2},\frac{N-1}{2},\frac{N-1}{2}}_{[N+1]}.\n
	\end{eqnarray}
Applying Eq. (\ref{eigen-back}) to
the above state, it becomes
	\begin{eqnarray}\label{product-1}
	\ket{\Psi_P(t)}&=&
	\frac{\sqrt{N}(e^{-i\frac{N}{2}t}-e^{i\frac{N+2}{2}t})}{N+1}\ket{\Up}\ket{\frac{N}{2},\frac{N-1}{2},\frac{N-2}{2}}_{[N]}\n
	& &+\frac{e^{-i\frac{N}{2}t}+Ne^{i\frac{N+2}{2}t}}{N+1}\ket{\Down}\ket{\frac{N}{2},\frac{N-1}{2},\frac{N}{2}}_{[N]},
	\end{eqnarray}
which allows us to directly calculate the reduced density matrix for the central spin, i.e.,
\begin{eqnarray}\label{11}
\rho_0^P(t)=a_{11}(t)\ket{\Up}\bra{\Up}+(1-a_{11}(t))\ket{\Down}\bra{\Down},
\end{eqnarray}
with $a_{11}(t)=2N[1-\cos((N+1)t)]/(N+1)^2$.
By definition (\ref{fidelity}), the evolution of fidelity is obtained from (\ref{11}) as follows:
\begin{equation}\label{F-product}
F_0^P(t)=1-a_{11}(t)=1-\frac{2N}{(N+1)^2}[1-\cos((N+1)t)].
\end{equation} 

This simple form (\ref{F-product}) provides rich insights into the central spin decoherence problem.
The dynamic term in Eq. (\ref{F-product}) describes an oscillation with no long-time decay where the frequency scales as $\mathcal O(N)$ and the amplitude scales as $\mathcal O(1/N)$ in the thermodynamic limit $N\to\infty$, which has been pointed out in Ref. \cite{Bortz2} by calculating $\ex{S_0^z(t)}$.
Moreover, the $\mathcal O(1/N)$-oscillation of  fidelity indicates that a strong magnetic field can suppress decoherence by polarizing a large number of bath spins \cite{Hanson,Burkard}.
The absence of a long-time decay can be understood  from the energy differences of eigenstates that  determine transition frequencies.
In homogeneous CSM, the distribution of gaps $g_i = E_{i+1}-E_i$ of adjacent energies is almost uniform, see Eq. (\ref{eigen-solution}), and thus, the fidelity evolution (\ref{F-product}) displays a periodic behavior even for more complex initial state settings (\ref{with-entangled-pair}, \ref{EP-C}). 
When the couplings between the central spin and the bath spins become inhomogeneous, the distribution of $g_i$ is no longer uniform leading to a long-time decay of $\ex{S_0^z(t)}$ \cite{Khaetskii1,Khaetskii} but  $\ex{S_0^z(t)}$ will not tend to a stable value.  
Eventually, $\ex{S_0^z(t)}$ for such a fully polarized bath reaches a persistent oscillation with an amplitude of $\mathcal O(1/N)$ \cite{Bortz2}.
On the other hand, if the CSM with a disorder magnetic field for bath spins instead of inhomogeneous couplings, a phase transition will occur from the eigenstate thermalization hypothesis (ETH) phase  to the many-body localization (MBL) phase when disorder  overs interaction.
Such phenomenon has been also witnessed by level statistics $\ex{\min(g_i,g_{i+1})/\max(g_i,g_{i+1})}_i$ \cite{Yao} and a long-time decay will occur \cite{Serbyn,Zhou}.
The difference from the inhomogeneous case is that the oscillation of $\ex{S_0^z(t)}$ decays completely to a constant in the CSM with a disorder field. 
Based on the above observations, 
non-uniformity of level statistics caused by disorder magnetic fields or inhomogeneous couplings takes main responsibility for the emergence of long-time decays.
Considering that there is no long-time decay in our model (\ref{Hamiltonian}), we will  use the scaling of leading oscillation amplitude of fidelity evolution to characterize central spin decoherence and call the scaling of leading oscillation amplitude of $X$ the \emph{amplitude scaling } of $X$ for convenience.

\subsection{The effects of entanglement}
Now we try to seek a decoherence suppression beyond $\mathcal O(1/N)$ by considering entangled baths, namely the GHZ-bath:
$\ket{\rm GHZ}_{[N]}=( \ket{\up}^{\otimes N}+\ket{\downarrow}^{\otimes N} )/\sqrt{2}$ and 
the $W$-bath \cite{W}: $\ket{W}_{[N]}=\left(\ket{\downarrow\uparrow\cdots\uparrow}+\cdots+\ket{\uparrow\cdots\uparrow\downarrow}\right)/\sqrt{N}$.
And the initial states are given by
	\begin{eqnarray}
	\ket{\Psi_{GHZ} (0)}&=&\ket{\Downarrow}\ket{GHZ}_{[N]}\n
	&=& \frac{\ket{\Down}}{\sqrt{2}}\lf(\ket{\frac{N}{2},\frac{N-1}{2},\frac{N}{2}}_{[N]}+\ket{\frac{N}{2},\frac{N-1}{2},-\frac{N}{2}}_{[N]}\rg)\label{GHZ-bath}\n
	\label{W-bath}
	\ket{\Psi_{W} (0)}&=&\ket{\Downarrow}\ket{W}_{[N]}\n
	&=&\ket{\Down}\ket{\frac{N}{2},\frac{N-1}{2},\frac{N-2}{2}}_{[N]}.
	\end{eqnarray}
In other words, we only consider the effect of entanglement among baths.
Similarly, we use Eq. (\ref{eigen1}) to determine the states after a unitary time evolution,
	\begin{eqnarray}
	\ket{\Psi_{GHZ}(t)}
	&=&\frac{1}{\sqrt{2(N+1)}}e^{-i\frac{N}{2}t}\ket{\frac{N+1}{2},\frac{N}{2},\frac{N-1}{2},\frac{N-1}{2}}_{[N+1]}\n
	& &-\sqrt{\frac{N}{2(N+1)}}e^{i\frac{N+2}{2}t}\ket{\frac{N-1}{2},\frac{N}{2},\frac{N-1}{2},\frac{N-1}{2}}_{[N+1]}\n
	& &+\frac{1}{\sqrt{2}}e^{-i\frac{N}{2}t}
	\ket{\frac{N+1}{2},\frac{N}{2},\frac{N-1}{2},-\frac{N+1}{2}}_{[N+1]},\n
	\ket{\Psi_{W}(t)}
	&=&\sqrt{\frac{2}{N+1}}e^{-i\frac{N}{2}t}
	\ket{\frac{N+1}{2},\frac{N}{2},\frac{N-1}{2},\frac{N-3}{2}}_{[N+1]}\n
	& &-\sqrt{\frac{N-1}{N+1}}e^{i\frac{N+2}{2}t}\ket{\frac{N-1}{2},\frac{N}{2},\frac{N-1}{2},\frac{N-3}{2}}_{[N+1]}.
	\end{eqnarray}
With these explicit expressions of reduced density matrices, it is effortless to calculate  fidelity for the central spin.
The corresponding results are as follows
\begin{eqnarray}\label{F-GHZ}
&&F_0^{GHZ}(t)=1-\frac{N}{(N+1)^2}[1- \cos((N+1)t)],\\
&&F_0^{W}(t)=1-\frac{4(N-1)}{(N+1)^2}[1-\cos((N+1)t)].\label{F-W}
\end{eqnarray}
It is observed from Eqs. (\ref{F-GHZ}) and (\ref{F-W}) that the amplitude scalings of fidelity for the GHZ-bath and the $W$-bath are both $\mathcal O(1/N)$ although  they belong to distinct entanglement classes in the entanglement classification problem  \cite{Bastin}.
The same amplitude scaling of fidelity for the entangled baths and the product bath (\ref{F-product}) indicates that such entangled baths (\ref{GHZ-bath}) cannot provide a more effective decoherence suppression. 
One possible reason for this phenomenon is that the initial states we have been considered so far are all in a product form, i.e, $\ket{\Down}\otimes\ket{\rm bath}$, lacking of system-bath entanglement, which causes the failure of entanglement to affect the dynamics of the central spin in our model.

Here, we start to construct a maximal entangled pair between the central spin and the first bath spin as the initial state,
\begin{equation}\label{with-entangled-pair}
\ket{\Psi_{EP}(0)}=\frac{1}{\sqrt{2}}(\ket{\Downarrow}\ket{\uparrow}+\ket{\Uparrow}\ket{\downarrow})\ket{\uparrow\uparrow\ldots\uparrow}_{[N-1]}.
\end{equation}
As before, we use Eqs. (\ref{eigen1}) and (\ref{eigen-back}) to derive
the fidelity of the central spin,
\begin{eqnarray}\label{F-EP}
F_0^{EP}(t)&=&\frac{1}{2}+\sqrt{\frac{1}{4}-\left\{\frac{2(N-1)[\cos((N+1)t)-1]}{(N+1)^2}\right\} ^2}.
\end{eqnarray}
A direct simplification of Eq. ($\ref{F-EP})$ gives the amplitude scaling of  fidelity being $\mathcal O(1/N^2)$.
In this sense, the state with a system-bath entangled pair (\ref{with-entangled-pair}) outperforms than the product state (\ref{Product-bath}) in suppressing of central spin decoherence.

\section{Entanglement enhances coherence}\label{Example}
In the previous section,  fidelity is utilized to quantify  decoherence of the central spin. 
It is worth noting that the reduced density matrices of the central spin $\rho_0(t)$ discussed before do not contain off-diagonal elements ($\ket{\Up}\bra{\Down}$, $\ket{\Down}\bra{\Up})$ during  time evolutions and thus only the effect of amplitude decoherence is involved.
In this section, we fix our attention on the phase decoherence problem and aim to improve the quantum coherence of the central spin encoded in the off-diagonal elements of its density matrix.
The relative entropy of coherence   $C_0(t)$ (\ref{relative-entropy}) is applied in here to characterize such quantum coherence since $C_0(t)$ is nonzero if and only if the reduced density matrix of the central spin has  non-diagonal elements.

\begin{figure}[ht!]
	\centering
	\includegraphics[width=1.0\linewidth]{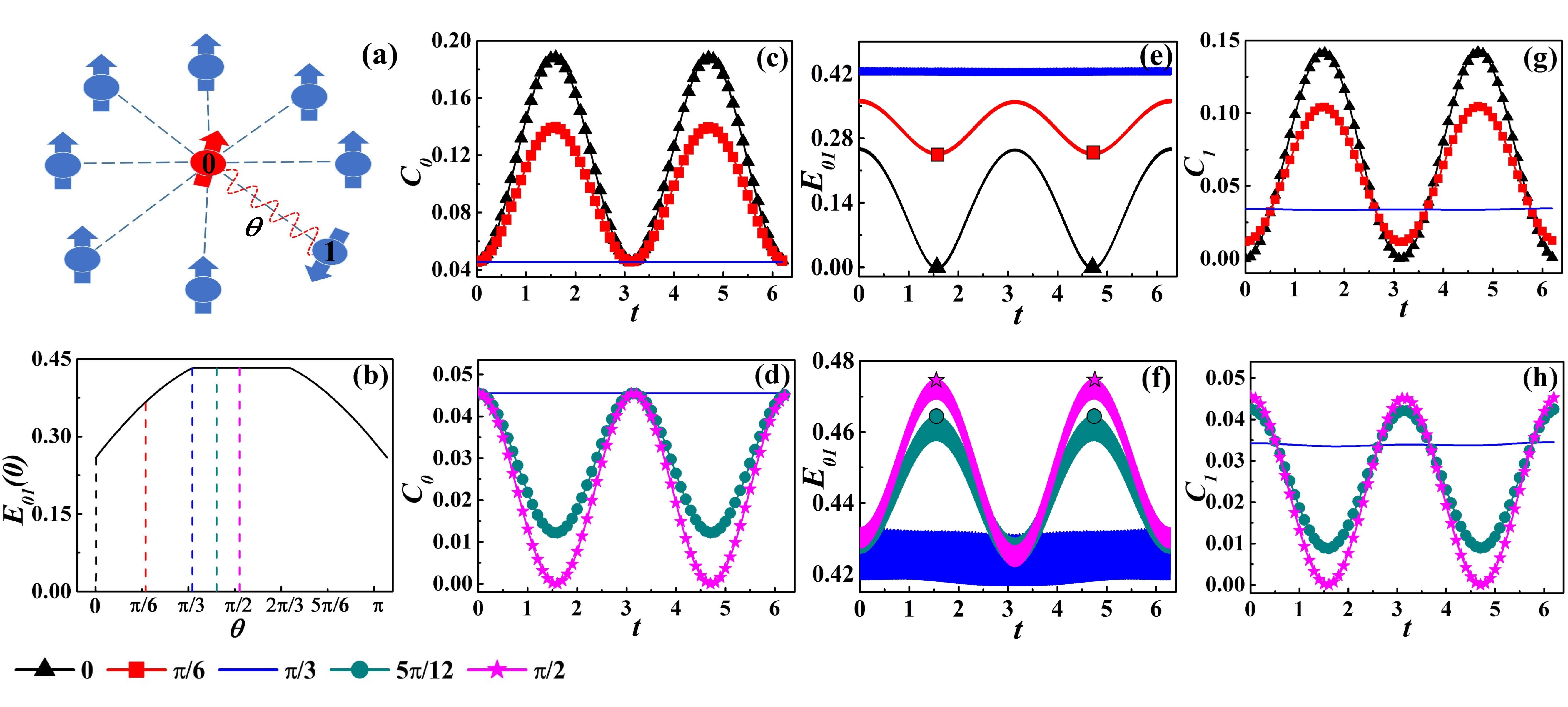}
	\caption{
		a The initial states with an entangled pair between the central spin  and the first bath spin  denoted  by an red curve line.
		b By changing the value of $\theta$, we can set different  initial entanglement values $E_{01}(0)$ for the entangled pair.  
		During the evolution, the coherence values of the central spin $C_0(t)$ (c) and the first bath spin $C_1(t)$ (g) both excess their initial values, while the concurrence of entangled pair $E_{01}(t)$ behaves just opposite (e). In addition, when $\pi/3\leq\theta<\pi/2$, the coherence of $C_0(t)$ (d) and $C_1(t)$ (h) are less than their initial values and the opposite behaves also holds with $E_{01}(t)$ (f).
		The number of bath spins $N$ is set to 500.}
	\label{fig:CS}
\end{figure}

In section (\ref{decoherence-problem}), initial system-bath entanglement helps the central spin to prevent its amplitude decoherence. Now we expect to improve the quantum coherence of the central spin during the dynamics with the help of some  system-bath entangled pair. 
Then, we consider a set of initial states as follows:
\begin{eqnarray}\label{EP-C}
\rho(\theta)=\rho^{EP}_{[2]}(\theta)\otimes\rho_{[N-1]}^{bath},
\end{eqnarray}
where the central spin and the first bath spin constitute an entangled pair $\rho^{EP}_{[2]}$, while all bath spins except the first bath spin constitute a $(N-1)$-qubit product state $\rho_{[N-1]}^{bath}=\ket{\up\up\cdots\up}\bra{\up\up\cdots\up}_{[N-1]}$.
The parameter $\theta$ is required to adjust the bipartite entanglement of the entangled pair but, at the same time, to keep the coherence of the central spin unchanged.
It was  proved in Ref. \cite{Wharton} that an arbitrary non-maximal entangled two-qubit pure state $\ket{\phi}_{01}=a\ket{\Up\up}+b\ket{\Up\down}+c\ket{\Down\up}+d\ket{\Down\down}$ can be parameterized in terms of six angles $(\chi,\theta_0,\phi_0,\theta_1,\phi_1,\gamma)$:
	\begin{eqnarray}\label{para}
	&&a=\lf[\cos\frac{\chi}{2}\cos\frac{\theta_0}{2}\cos\frac{\theta_1}{2}e^{i\frac{\gamma}{2}}\!+\!\sin\frac{\chi}{2}\sin\frac{\theta_0}{2}\sin\frac{\theta_1}{2}e^{-i\frac{\gamma}{2}}\rg]e^{-i\frac{\phi_0+\phi_1}{2}},\n
	&&b=\lf[\cos\frac{\chi}{2}\cos\frac{\theta_0}{2}\sin\frac{\theta_1}{2}e^{i\frac{\gamma}{2}}\!-\!\sin\frac{\chi}{2}\sin\frac{\theta_0}{2}\cos\frac{\theta_1}{2}e^{-i\frac{\gamma}{2}}\rg]e^{-i\frac{\phi_0-\phi_1}{2}},\n
	&&c=\lf[\cos\frac{\chi}{2}\sin\frac{\theta_0}{2}\cos\frac{\theta_1}{2}e^{i\frac{\gamma}{2}}\!-\!\sin\frac{\chi}{2}\cos\frac{\theta_0}{2}\sin\frac{\theta_1}{2}e^{-i\frac{\gamma}{2}}\rg]e^{i\frac{\phi_0-\phi_1}{2}},\n
	&&d=\lf[\cos\frac{\chi}{2}\sin\frac{\theta_0}{2}\sin\frac{\theta_1}{2}e^{i\frac{\gamma}{2}}\!+\!\sin\frac{\chi}{2}\cos\frac{\theta_0}{2}\cos\frac{\theta_1}{2}e^{-i\frac{\gamma}{2}}\rg]e^{i\frac{\phi_0+\phi_1}{2}}.\nonumber
	\end{eqnarray}
This parameterization has a geometric intuition.
For instance, the reduced density matrix of the central spin can be expressed as
\begin{eqnarray}\label{bloch sphere}
\rho_0=\Tr_1(\ket{\phi}\bra{\phi})=\frac{I+ \mathbf{r_0}\cdot\bm{\sigma}}{2},
\end{eqnarray}
with $\mathbf{r_0}=(\cos\chi,\theta_0,\phi_0)$ in spherical coordinates 
and the reduced density matrix of the first bath spin is in the same form (\ref{bloch sphere}) with $\mathbf{r_1}=(\cos\chi,\theta_1,\phi_1)$. 
The parameter $\chi$ is not only related to the norms of the Bloch vectors $\mathbf{r_0}$ and $\mathbf{r}_1$, but also determines the value of the concurrence by $E_{01}(\ket{\phi})=\sin\chi$. 
Note that this parameterization (\ref{para}) excludes the maximal entangled two-qubit pure state, i.e., $\chi\neq \pi/2$, since for $\chi=\pi/2$ the norms of the Bloch vectors $|\mathbf{r_0}|$ and $|\mathbf{r_1}|$ both are zero, which is absurd.
Then we take $\theta_1$ as the parameter $\theta$ in Eq. (\ref{EP-C}) and fix the other angles $(\chi,\theta_0,\phi_0,\phi_1,\gamma)$ to $(\pi/3,\pi/2,0,0,0)$ as an example.
In this setting, the reduced density matrix of the central spin is a constant matrix and thus the  coherence of the central spin no longer depends on the parameter $\theta$.
However, at this time, the concurrence is also constant due to $\chi$ being fixed.
According to definition (\ref{concurrence2}), concurrence depends on the optimal ensemble decomposition of a given density matrix.
The unique ensemble decomposition of a quantum state $\rho$ up to a phase factor exists only when $\rho$ is a pure state.
Thus, we mix the state $\ket{\phi}\bra{\phi}$ and $(\ket{\Up\up}\bra{\Up\up}+\ket{\Down\down}\bra{\Down\down})/2$ with  equal possibility $1/2$ to construct a set of entangled pair,
\begin{eqnarray}\label{EP}
\rho_{[2]}^{EP}(\theta)=(\frac{1}{2}\ket{\phi}\bra{\phi}+\frac{1}{4}\ket{\Uparrow\uparrow}\bra{\Uparrow\uparrow}+\frac{1}{4}\ket{\Downarrow\downarrow}\bra{\Downarrow\downarrow}),
\end{eqnarray} 
and expect their optimal decompositions to be different for  different $\theta$.
Here, 
\begin{eqnarray}
&&\ket{\phi}=a\ket{\Uparrow\uparrow}+b\ket{\Uparrow\downarrow}+c\ket{\Downarrow\uparrow}+d\ket{\Downarrow\downarrow},\\
&&a=\frac{\sqrt{6}}{4}\cos\frac{\theta}{2}+\frac{\sqrt{2}}{4}\sin\frac{\theta}{2},\,
b=\frac{\sqrt{6}}{4}\sin\frac{\theta}{2}-\frac{\sqrt{2}}{4}\cos\frac{\theta}{2},\n
&&c=\frac{\sqrt{6}}{4}\cos\frac{\theta}{2}-\frac{\sqrt{2}}{4}\sin\frac{\theta}{2},\,
d=\frac{\sqrt{6}}{4}\sin\frac{\theta}{2}+\frac{\sqrt{2}}{4}\cos\frac{\theta}{2}.\nonumber
\end{eqnarray}
For the entangled pair (\ref{EP}),  the reduced density matrix of the central spin reads
\begin{eqnarray}
\rho_0(\theta)=\frac{2I+\mathbf{r_0}\cdot\bm{\sigma}}{4},
\end{eqnarray}  
with $\mathbf{r}_0=(1/2,\pi/2,0)$ in spherical coordinates and  the value of coherence is a constant of $[5(\log_25)/8+3(\log_23)/8-2]\simeq 0.0456$ according to Eq. (\ref{relative-entropy}).
In Fig. \ref{fig:CS}b  we plot the concurrence of  the entangled pair (\ref{EP}) versus $\theta$ where the concurrence first increases, then remains constant, and finally decreases to the initial value.

Having a set of initial states (\ref{EP-C}) with the same initial coherence value for the central spin but different initial system-bath entanglement values,
we are going to investigate their dynamics.
The explicit expression of $\rho_{01}(t;\theta)=\Tr_{2,3,\ldots,N}(e^{-iHt}\rho(\theta)e^{iHt})$ is found in "Appendix". 
Omitting $\mathcal O(1/N)$ terms in $\rho_{01}(t;\theta)$, the reduced density matrix of the central spin is simplified to
\begin{eqnarray}
\rho_{0}(t;\theta)\simeq
\begin{bmatrix}
A_{11}(t)&& A_{12}(t)\\
A_{12}^*(t)&& 1-A_{11}(t)
\end{bmatrix},
\end{eqnarray}
with $A_{11}(t)\simeq1/2$ and $A_{12}(t)\simeq1/8+ac(e^{-it(N+1)}-1)/2+bd(e^{-it(N-1)}-1)/2.$
It follows that 
the evolution of central spin coherence is
\begin{eqnarray}\label{C_0}
C_0 (t;\theta)\simeq1-H_b(\lambda(t;\theta)),
\end{eqnarray}
where $H_b(x)=-x\log_2(x)-(1-x)\log_2(1-x)$ is the binary  entropy function and $\lambda(t;\theta)$ is one of the eigenvalues of $\rho_0(t;\theta)$, 
\begin{eqnarray}\label{cs-go}
\lambda(t;\theta)
=\frac{1}{2}+\frac{1}{8}\sqrt{1+\lf(\frac{1}{2}+\cos(2\theta)\rg)(1-\cos(2t))}.
\end{eqnarray}
Notice that $H_b(x)=H_b(1-x)$ and $H_b(x)$ is a monotonically decreasing function of $x$ when $1/2\leq x\leq1$.
According to Eqs. (\ref{C_0}) and (\ref{cs-go}), we observe that when $0\leq\theta<\pi/3$, i.e., $[1/2+\cos(2\theta)]>0$, the coherence of the central spin $C_0(t;\theta)$  first increases and then declines to its initial value in a time period (Fig. \ref{fig:CS}c), 
while $C_0(t)$ behaves just opposite as $\pi/3<\theta\leq\pi/2$ (Fig. \ref{fig:CS}d).
Therefore, for the system-bath entangled pair $\rho_{[2]}^{EP}(\theta)$ with the condition of $0\leq\theta<\pi/3$, the central spin coherence will increase over time.

To shed light on the origin of such dynamics behaviors, the concurrence of entangled pair is plotted in Fig.\ref{fig:CS}e-f by means of the reduced density matrices (\ref{appendix}) given in "Appendix".
As seen from Fig.\ref{fig:CS}c-f,
the concurrence $E_{01}(t)$ will decrease at the region of central spin $C_0(t)$ increasing, vice versa.
Thus, in the case of $0\leq\theta<\pi/3$, the  increase in central spin coherence is derived from the loss of entanglement in the entangled pair.
The initial coherence is, however, nonzero only for the central spin and the first bath spin.
It is necessary to calculate the evolution of coherence for the first bath spin $C_1(t)$ to exclude the possibility of coherence transfer between the central spin and the first bath spin leading to the gain of coherence in the central spin.

Similarly, we omit the $\mathcal O(1/N)$ terms of $\rho_{01}(t;\theta)$ (\ref{appendix}) in the thermodynamic limit and trace out its central spin degrees of freedom to obtain the evolution of the reduced density matrix for the first bath spin,
\begin{eqnarray}
\rho_1(t;\theta)\simeq
\begin{bmatrix}
1-B_{22}(t)&& B_{12}(t)\\
B_{12}^*(t)&& B_{22}(t)
\end{bmatrix},
\end{eqnarray}
with $B_{22}(t)\simeq1/4+d^2/2+b^2/2$ and $B_{12}(t)=ab\exp(-it)/2+cd\exp(it)/2$.
By definition (\ref{relative-entropy}), the coherence of the first bath spin is given by 
\begin{eqnarray}\label{C1}
C_1(t;\theta)&\simeq&H_b\lf(\frac{1}{2}+\frac{\cos\theta}{8}\rg)
-H_b(\lambda'(t;\theta)),
\end{eqnarray}
where 
\begin{eqnarray}\label{lambed'}
\lambda'(t;\theta)=\frac{1}{2}\!+\!\frac{1}{8}\sqrt{\frac{3}{2}+\cos(2\theta)-\lf(\frac{1}{2}+\cos(2\theta)\rg)\cos(2t)}.\n
\end{eqnarray}
It is obvious from  Eqs. (\ref{C_0})-(\ref{cs-go}) and  (\ref{C1}-\ref{lambed'}) that the monotonicity of $C_1(t)$ and $C_0(t)$ is the same,
which confirms that the consumption of entanglement in this entangled pair is the main source of coherence gains in the central spin as $0\leq\theta<\pi/3$, and illustrated in Figs.\ref{fig:CS}e-h.
Using the system-bath entangled pair, 
we realize quantum coherence-enhanced dynamics for the central spin by using of the entanglement between bath baths and the central spin.

\section{Conclusions}\label{4}
We have investigated the effects of entanglement in the central spin decoherence problem, amplitude decoherence, and phase decoherence, by obtaining exact evolutions of fidelity, concurrence, and the relative entropy of coherence.
The closed-form expressions of them have been obtained in this paper and in the thermodynamic limit we extracted their amplitude scalings summarized in Table. \ref{scalings}.
\begin{table}[ht!]
	\begin{center}
	\caption{\label{scalings}Amplitude scalings of four different initial states}
	\begin{tabular}{lcccc}\hline
		\hline
		Scalings&Product bath&$W$-bath&GHZ-bath&Entangled pair\\\hline
		$F_0(t)$ &$\mathcal O(1/N)$&$\mathcal O(1/N)$&$\mathcal O(1/N)$&$\mathcal O(1/N^2)$\\
		\hline
		\hline
	\end{tabular}
    \end{center}
\end{table}
Here, the degree of amplitude decoherence for the central spin is characterized by the amplitude scaling of fidelity $F_0(t)$. 
For the initial states with only bath entanglement, the entangled bath ($W$-bath or GHZ-bath) dose not outperform than the product bath in suppression of central spin decoherence.
In contrast to the above situation, however, the state with a maximal entangled pair establishes an $\mathcal {O}(1/N^2)$  suppression of amplitude decoherence.

Since quantum coherence is embedded in the off-diagonal elements of the density matrix, the relative entropy of coherence has been used to describe quantum coherence dynamics for the central spin and to analyze its phase decoherence problem.
We provide a method to construct some initial states with suitable entangled pair where the quantum coherence of the central spin can be improved over its initial value in dynamics.
By calculating the evolutions of central spin coherence and concurrence of the entangled pair, we confirmed that this part of increased coherence comes from the consumption of the entanglement in entangled pair.

The central spin model, as an ideal model to characterize the quantum dot systems, has been extensively studied.
In quantum dot system, the bath spins are unable to manipulate.
However, the rapid development of quantum technologies  enables us to precisely manipulate individual qubits and to make use of the unique quantum properties, such as superposition and entanglement. 
For example, multi-component atomic Schr\"odinger cat states have been prepared up to 20 qubits in superconductor system \cite{hang f}.
In the nitrogen-vacancy center, control of 10 qubits  has been realized \cite{Bradley} and two-qubit entanglement can be preserved for over 10 seconds.
The tremendous progress has been made in these quantum computing platforms to make it possible to realize the states we discussed.
Moreover, the central spin structure is realizing its potential in developing quantum technologies, e.g., quantum memory  \cite{poten}. 
Our research sheds light to further investigate the effects of initial system-bath entanglement in more complicated central spin models and to apply central spin models to some quantum technologies.

\section{Acknowledgements}
We thank Xi-Wen Guan whose comments and	suggestions have been of great help in this work.
This work was supported by NSFC (Grants Nos. 11975183 and 11875220), the Key Innovative Research Team of Quantum Manybody theory and Quantum Control in Shaanxi Province (Grant No. 2017KCT-12), the Major Basic Research Program of Natural Science of Shaanxi Province (Grant No. 2017ZDJC-32), the National Level College Students Innovation and Entrepreneurship Training Program in Northwest University (Grant Nos. 201910697012 and 201910697121), and the Double First-Class University Construction Project of Northwest University.

\section*{Appendix}\label{app}
	The reduced density matrix of the entangled pair (\ref{EP}) is given by
	\begin{eqnarray}\label{appendix}
	\rho_{01}(t;\theta)&=&
	\begin{bmatrix}
	D_{11}(t)&&D_{12}(t)&&D_{13}(t)&&D_{14}(t)\\
	D_{12}^*(t)&&D_{22}(t)&&D_{23}(t)&&D_{24}(t)\\
	D_{13}^*(t)&&D_{23}^*(t)&&1-D_{11}(t)-D_{22}(t)-D_{44}(t)&&D_{34}(t)\\
	D_{14}^*(t)&&D_{24}^*(t)&&D_{34}^*(t)&&D_{44}(t)
	\end{bmatrix},
	\end{eqnarray}
	where the matrix elements are
	\begin{small}
		\begin{eqnarray}
		D_{11}(t)&=&
		\frac{1}{4}+\frac{a^2}{2}
		-\frac{(1+2d^2)(N-2)(\cos[(N-1)t]-1)}{N^2(N-1)}
		-\frac{2(1+2d^2)(N-2)}{(N+1)N^2}(\cos t -\cos(Nt))\n
		& &+\frac{N-1}{(N+1)^2N}\left\lbrace -\frac{N-2}N+(b+c)(b-cN)
		-\frac{2d^2(N-2)}{N} \right\rbrace\lf[\cos [(N+1)t]-1\rg]\n
		& &+\frac{-b(b+c)(N-1)}{N(N+1)}(\cos t-1)
		+\frac{-b(N-1)(b-cN)}{(N+1)N^2}(\cos(Nt)-1),\n
		D_{22}(t)&=&
		\frac{b^2}{2}-\frac{2+d^2(N-2)^2}{N^2(N-1)}\lf\lbrace\cos[(N-1)t]-1\rg\rbrace
		+\frac{\cos t-1}{N(N+1)}\lf(\frac{2(N-2)}{N}+\frac{4d^2(N-2)}{N}+b(b+c)(N-1)\rg)\n
		& &+\frac{\cos(Nt)-1}{(N+1)N^2}\lf(-2(N-2)-4d^2(N-2)+b(b-cN)(N-1)\rg)\n
		& &+\frac{1}{(N+1)^2N}\left\lbrace \frac{-2(N-1)}{N}+(b+c)(b-cN)-\frac{4d^2(N-1)}{N}\right\rbrace [\cos[(N+1)t]-1],\n
		D_{44}(t)&=&
		\frac{2d^2+1}{4(N+1)^2N^2}\lf\lbrace 4(N-1)\cos[(N+1)t]+[16(N+1)-10(N+1)^2+2(N+1)^3]\cos t\rg.\n
		& &\lf.+[6(N+1)^2-16(N+1)]\cos(Nt)+[2(N+1)^3-6(N+1)^2]\cos((N-1)t)\rg.\n
		& &\lf.+8-4(N+1)+11(N+1)^2-6(N+1)^3+(N+1)^4\rg\rbrace,\n
		D_{12}(t)&=&
		\frac{ab}{2}+\frac{ab(N-1)}{2N}(e^{-it}-1)+\frac{a(b-cN)}{2N(N+1)}\lf(e^{-it(N+1)}-1\rg)\n
		& &+\frac{1}{2N(N+1)(N-1)}\lf\lbrace\frac{2bd(N-1)^2}{N}\lf(e^{-itN}-1\rg)+\frac{bd(N+1)(N-1)(N-2)}{N}\lf( e^{-it(N-1)}-1\rg)\rg.\n
		& &\lf.-d(b+c)(N-1)(N-2)\lf(e^{-itN}-1\rg)-\frac{2d(b+c)(N-1)^2}{N+1}\lf(e^{-it(N+1)}-1\rg)-\frac{2bd(N-1)^2}{N}(e^{it}-1)\rg.\n
		& &\lf.+\frac{d(2b-2cN)(N-1)^2}{N(N+1)}\lf(e^{it(N+1)}-1\rg)
		+\frac{d(N-1)(N-2)(b-cN)}{N}\lf(e^{itN}-1\rg)\rg.\n
		& &\lf.+\lf(d(b+c)(N-1)(N-2)-\frac{d(N-1)(N-2)(b-cN)}{N}\rg)(e^{-it}-1)    \rg\rbrace,\n
		D_{13}(t)&=&
		\frac{ac}{2}
		-\frac{bd(N-1)^2+d(b+c)(N-2)N}{2(N+1)N^2}\lf( e^{-itN}-1\rg)
		+\frac{bd(N-2)}{2N^2}\lf(e^{-it(N-1)}-1\rg)
		-\frac{bd(N-1)}{(N+1)N^2}\lf(e^{it}-1\rg)\n
		& &-\frac{a(b-cN)N(N+1)-d(b+c)(N-1)^2}{2(N+1)^2N}\lf(e^{-it(N+1)}-1\rg)
		-\frac{d(b-cN)}{2(N+1)N^2}\lf(e^{itN}-1\rg)\n
		& &-\frac{d(b+c)}{2N(N+1)}\lf(e^{-it}-1\rg)
		+\frac{d(N-1)(b-cN)}{(N+1)^2N^2}\lf(e^{it(N+1)}-1\rg)
		-\frac{d(b-cN)(N-2)}{2(N+1)N^2}\lf(e^{it}-1\rg),\n
		D_{14}(t)&=&
		\frac{ad}{2}
		+\frac{ad}{2N}\lf(e^{-it}-1\rg)
		+\frac{ad(N-2)}{2N}\lf(e^{-itN}-1\rg)
		+\frac{ad(N-1)}{2N(N+1)}\lf(e^{-it(N+1)}-1\rg) ,\n
		D_{23}(t)&=&
		\frac{bc}{2}
		-\frac{d^2}{2}\left\lbrace \frac{2+(N-1)(N-2)}{(N+1)N^2}\lf(e^{-it}-1\rg)
		+\frac{2(N-3)}{(N+1)N^2}\lf(e^{itN}-1\rg)
		+\frac{4(N-1)}{(N+1)^2N^2}\lf(e^{it(N+1)}-1\rg)\rg.\n
		& &\lf.-\frac{2(N-1)^2}{N^2(N+1)^2}\lf(e^{-it(N+1)}-1\rg)
		+\frac{2(N-2)}{(N+1)N^2}\lf(e^{-itN}-1\rg)
		-\frac{(N-1)(N-2)}{(N+1)N^2}\lf(e^{-itN}-1\rg)\right\rbrace \n
		& &+\frac{-(b+c)(b-cN)N^2+(N-1)^2}{2(N+1)^2N^2}\lf(e^{-it(N+1)}-1\rg)
		+\lf(\frac{-b(N-1)(b-cN)}{2N(N+1)}+\frac{(N-2)(N-3)}{4(N+1)N^2}\rg)\lf(e^{-itN}-1\rg)\n
		& &+\lf(1+2d^2\rg)\lf(\frac{N-2}{4N^2(N-1)}\lf(e^{it(N-1)}-1\rg)
		+\frac{N-2}{(N+1)N^2}\lf(e^{it}-1\rg)-\frac{(N-2)^2}{4N^2(N-1)}\lf(e^{-it(N-1)}-1\rg)\rg)\n
		& &-\frac{N-1}{(N+1)^2N^2}\lf(e^{it(N+1)}-1\rg)
		-\frac{N-3}{2(N+1)N^2}\lf(e^{itN}-1\rg)
		-\frac{(N+1)^2-5(N+1)+8}{4(N+1)N^2}\lf(e^{-it}-1\rg)\n
		& &+\frac{(b+c)(b-cN)}{2(N+1)^2N}\lf(e^{it(N+1)}-1\rg)
		+\frac{b(b+c)(N-1)}{2N(N+1)}\lf(e^{it}-1\rg),\n
		D_{24}(t)&=&
		\frac{bd}{2}
		+\frac{d(b+c)}{2N(N+1)}\lf(e^{-it}-1\rg)
		+\frac{d(b-cN)}{(N+1)^2N^2}\lf(e^{it(N+1)}-1\rg)
		+\frac{d(b-cN)}{2(N+1)N^2}\lf(e^{itN}-1\rg)\n
		& &+\lf(\frac{bd(N-1)^2}{2(N+1)N^2}+\frac{d(b+c)(N-2)}{2N(N+1)}\rg) \lf(e^{-itN}-1\rg)
		+\frac{d(b+c)(N-1)}{2(N+1)^2N}\lf(e^{-it(N+1)}-1\rg)\n
		& &+\frac{2bd(N-1)+d(b-cN)(N-2)}{2(N+1)N^2}\lf(e^{it}-1\rg)
		+\frac{bd(N-1)(N-2)}{2N^2}\lf(e^{-it(N-1)}-1\rg),\n
		D_{34}(t)&=&
		\frac{cd}{2}
		+\frac{d(b+c)}{2N(N+1)}\lf(e^{-it}-1\rg)
		-\frac{d(b-cN)}{(N+1)^2N}\lf(e^{it(N+1)}-1\rg)
		-\frac{d(b-cN)}{2N(N+1)}\lf(e^{itN}-1\rg)\n
		& &+\frac{d(b+c)(N-1)}{2(N+1)^2N}\lf(e^{-it(N+1)}-1\rg)
		+\frac{d(b+c)(N-2)}{2N(N+1)}\lf(e^{-itN}-1\rg)
		-\frac{d(b-cN)(N-2)}{2N(N+1)}\lf(e^{it}-1\rg).
		\end{eqnarray}
	\end{small}Similarly, the reduced density matrix of the second bath spin can be written as follows:
\begin{eqnarray}
\rho_2(t;\theta)&=&\begin{bmatrix}
1-E_{22}(t)&&E_{12}(t)\\
E_{12}^*(t)&&E_{22}(t)
\end{bmatrix},
\end{eqnarray}
and here the matrix elements are
\begin{small}
\begin{eqnarray}
    E_{12}(t)&=&\frac{-bd}{(N+1)N^2}\lf(e^{-itN}-1\rg)
    +\frac{1}{2(N+1)N}\lf((b-cN)\lf(a-\frac{2d}{N+1}\rg)-\frac{d(b+c)(N-1)}{N+1}\rg)\lf(e^{-it(N+1)}-1\rg)\n
    &&+\frac{1}{2(N+1)N^2(N-1)}[d(b-cN)(N-2)(N+1)+2bd(N-1)]\lf(e^{it}-1\rg)
    -\lf(\frac{ab}{2N}+\frac{d(b+c)}{2(N+1)N}\rg)\lf(e^{-it}-1\rg)\n
    & &+\frac{d(b-cN)(N-1)}{(N+1)^2N^2}\lf(e^{it(N+1)}-1\rg)
    -\frac{bd(N+1)(N-2)}{2N^2(N-1)}\lf(e^{-it(N-1)}-1\rg)+\frac{d(b-cN)}{(N+1)N^2(N-1)}\lf(e^{itN}-1\rg),\n
	E_{22}(t)&=&\frac{1}{(N+1)N^2(N-1)}\lf(-\frac{1+2d^2}{2}(N^2-3N+4)-2(1+2d^2)(N-2)-b(b+c)N(N-1)\rg)(\cos t-1)\n
	&&+\frac{N-2}{N^2(N-1)^2}\lf(\frac{1}{2}-d^2(N+1)-\frac{2(N-1)}{N-2}\rg)(\cos[(N-1)t]-1)\n
	&&+\frac{1}{(N+1)N^2(N-1)}\lf((N-3)(d^2+\frac{1}{2})-b(b-cN)(N-1)\rg)(\cos(Nt)-1)\n
	&&+\frac{1}{(N+1)^2N^2(N-1)}\lf(-2(N-1)^2(d^2+\frac{1}{2})+(b+c)(b-cN)N(N-1)\rg)(\cos[(N+1)t]-1).	
\end{eqnarray}
\end{small}

\end{document}